\documentclass{jltp}
\usepackage{amssymb}
\usepackage{graphicx}

\runninghead{R. Blaauwgeers \textit{et al.}}{Quartz Tuning Fork}

\begin {document}
\title{Quartz Tuning Fork: Thermometer, Pressure- and Viscometer for Helium Liquids\thanks{Dedicated to former JLTP editor Frank Pobell}}
\author{R. Blaauwgeers$^1$, M. Blazkova$^2$, M. \v{C}love\v{c}ko$^3$, V.B. Eltsov$^{1,4}$\thanks{E-mail: ve@boojum.hut.fi} ,\\ R. de Graaf$^1$,
J. Hosio$^1$, M. Krusius$^1$, D. Schmoranzer$^5$,\\
  W. Schoepe$^6$,
  L. Skrbek$^{2,5}$, P. Skyba$^3$, R.E. Solntsev$^1$,\\ and D.E. Zmeev$^4$}
\address{$^1$Low Temperature Laboratory, Helsinki University of Technology,
  Espoo, Finland\\
$^2$Institute of Physics ASCR, Na Slovance 2,  182 21 Prague, Czech Republic\\
$^3$Centre of Low Temperature Physics,
Institute of Experimental Physics SAV,\\ Ko\v{s}ice, Slovakia\\
$^4$Kapitza Institute for Physical Problems, Kosygina 2, Moscow
119334, Russia\\
$^5$Faculty of Mathematics and Physics, Charles University,\\ Ke Karlovu 3, 121 16 Prague, Czech Republic\\
$^6$Fakult\"at f\"ur Physik, Universit\"at Regensburg, D-93040
Regensburg, Germany}
\date{\today}
\maketitle \vspace{-3mm}
\begin {abstract}

Commercial quartz oscillators of the tuning-fork type with a
resonant frequency of $\sim 32\,$kHz have been investigated in
helium liquids. The oscillators are found to have at best $Q$
values in the range $10^5$ --- $10^6$, when measured in vacuum
below 1.5\,K. However, the variability is large and for very low
temperature operation the sensor has to be preselected. We explore
their properties in the regime of linear viscous hydrodynamic
response in normal and superfluid $^3$He and $^4$He, by comparing
measurements to the hydrodynamic model of the sensor.

\end{abstract}
\vspace{-10mm}

\section{INTRODUCTION}

Quartz tuning forks are commercially produced piezoelectric
oscillators meant to be used as frequency standards in watches. An
extensive literature describes their use for a large number of
other additional applications.\cite{ForksGeneral} They have also
been employed in liquid He temperature measurements.\cite{Clubb}
This study was inspired by the expectation that industrially
produced quartz oscillators, with a calibrated standard frequency
of $2^{15}\,{\rm Hz}\; (= 32\,768\,$Hz) at room temperature, would
be reasonably identical and could be used as secondary
thermometers without need for recalibration.

We have performed measurements on four different forks of
identical dimensions, but produced by different manufacturers. It
turns out that without preselection at LHe temperatures the
results cannot be reduced on a common temperature dependence. For
instance, the resonance width $\Delta f_{\rm vac}$ measured in
vacuum below 1.5\,K proved to be 0.06\,Hz, 0.5\,Hz, 1.4\,Hz, and
0.06\,Hz for these four sensors. This measure of the intrinsic
dissipation, which limits the response of the device to its
environment at the lowest temperatures, cannot be determined from
room temperature measurements. It remains to be seen if simple
preselection criteria can be worked out, to narrow down the
variation in oscillator characteristics.

However, the quartz tuning fork offers other important advantages
as a sensor of its cryogenic environment. Forks are cheap and
readily available, they are robust and as such easy to install and
to use, they operate at a higher frequency than most other
vibrating sensors, and they are highly sensitive indicators of the
physical properties of the medium in which they are immersed. Thus
they provide handy \textit{in situ} information about the
conditions in a sample container at the far end of a
low-conductance filling line, which is helpful during flushing,
filling, emptying, and in general, for reproducible monitoring of
pressure and temperature changes. A major advantage in many
applications is that to drive these piezoelectric devices no
magnetic fields are needed and that they, in fact, are highly
insensitive to them.\cite{MagFields,Clubb}

In this report we explore the linear response of the quartz tuning
fork in He liquids at low excitation, in the regime of viscous
hydrodynamics. The purpose is to compare the measurements to a
hydrodynamic model which could explain the measured results. For
this Sec.~2 studies the oscillator properties of the tuning fork
in vacuum. In Sec.~3 the influence of the surrounding medium is
incorporated, Sec.~4 discusses briefly the practical measurement,
and later sections describe the measured results in vacuum as well
as in $^3$He and $^4$He liquids, by comparing the data to the
physical model. We postpone to a later occasion the analysis of
nonlinear effects and of the lowest temperatures with
collisionless motion of excitations. This latter aspect, the
creation and detection of excitations and of quantized vortices in
the $T \rightarrow 0$ temperature limit, is of great current
interest. A large amount of new information has been discovered on
vortex properties using vibrating resonators: (i)
spheres,\cite{Schoepe} grids,\cite{Skrbek} and wires\cite{Yano} in
$^4$He-II, (ii) grids\cite{Lancs} and wires\cite{Bradley} in
$^3$He-B, and (iii) wires\cite{Martikainen} in $^3$He--$^4$He
mixtures. One might hope that the quartz tuning fork could be used
for similar measurements.

\section{CHARACTERISTICS OF THE VIBRATING TUNING FORK}

\subsection{Mechanical Properties}

At sufficiently small oscillation amplitudes the fork can be
described as a harmonic oscillator subject to a harmonic driving
force $F = F_0 \cos(\omega t)$ and a drag force with linear
dependence on velocity. The equation of motion is given by
\begin{equation}
\frac{d^2x}{dt^2}+\gamma\frac{dx}{dt}+\frac{k}{m}x=\frac{F}{m}~.
\label{motion}
\end{equation}
We have here four parameters, namely the effective mass $m$ (of
one leg), the drag coefficient $\gamma$, the spring constant $k$,
and the amplitude of the driving force $F_0$. The effective mass
and the drag coefficient depend on the medium around the
oscillating fork.  The solution of this differential equation is
well known: It can be written as $x(t) = x_{\rm a}(\omega)
\sin(\omega t) + x_{\rm d}(\omega) \cos(\omega t) $, where $x_{\rm
a}$ and $x_{\rm d}$ are the absorption and dispersion,
respectively. The mean absorbed power $\langle F \, dx/dt\rangle =
F_0 \omega x_{\rm a}/2$ is at maximum at the resonant frequency
\[
\omega_0 = \sqrt\frac{k}{m}~.
\]
It is convenient to introduce the quality factor
\[
Q = \frac{\omega_0}{\gamma}
\]
as the ratio of the resonant frequency $\omega_0$ to the frequency
width $\Delta\omega = \gamma$, where $\Delta\omega$ is the full
width of the resonance curve at half of the maximum power.

\begin{figure}
\centerline{\includegraphics[width=2cm]{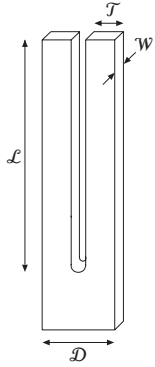}}
\caption{Sketch of the quartz tuning fork.} \label{forkgeom}
\end{figure}

The geometry of the fork is sketched in Fig.~\ref{forkgeom}. It is
characterized by the length $\cal L$, width $\cal W$, and
thickness $\cal T$ of a leg. The relevant vibration mode is the
basic antisymmetric mode, {\it i.e.} the one where the two legs of
the fork move in antiphase along the direction of $\mathcal{T}$.
Taking the known elasticity modulus $E$ of quartz, $E\,=
\,7.87\cdot10^{10}$ N/m$^2$, the spring constant is given
by\cite{Karrai1}
\[
k =
\frac{E}{4}\mathcal{W}\left(\frac{\mathcal{T}}{\mathcal{L}}\right)^3.
\]
and the effective mass of one leg in vacuum is
\begin{equation}
m_{\rm vac} = 0.24267\, \rho_{\rm q}\, \mathcal{L\,W\,T},
\label{mvac}
\end{equation}
where we use for the density $\rho_{\rm q}\,=\,2659$ kg/m$^3$, the
density of quartz (neglecting the electrodes on the legs).

We use forks with $\mathcal{L}=3.12$\,mm, $\mathcal{W}=0.352$\,mm,
$\mathcal{T}=0.402$\,mm, and $\mathcal{D}=1.0$\,mm. For this
geometry we find $k=1.48\cdot10^{4}$\,N/m and $m_{\rm
  vac} =$ $2.85\cdot10^{-7}$\,kg. This gives
$f_0=\omega_0/2\pi=36293$\,Hz, which is 11\% larger than the
manufactured value of 32768\,Hz at room temperature. The
discrepancy with the theoretical expression is most likely due to
additional weight of the evaporated electrodes, dependence of the
elasticity modulus of quartz on the orientation with respect to
the crystallographic axes, and deviations in geometry between the
real fork and the model.  At room temperature in vacuum the
measured width of the absorption curve at half power is typically
$\Delta f \approx 1.2$\,Hz, which gives $Q = f_0/\Delta f\approx
2.7\cdot10^4$. In room air the width increases to $\Delta f
\approx 4.3$\,Hz. With decreasing temperature the resonant
frequency diminishes and the $Q$ value increases. Our vacuum
measurements at LHe temperatures will be presented in
Sec.~\ref{vacuum}

\subsection{Electrical Properties and Calibration}
\label{ElectricalProperties}

The fork is excited with ac voltage $U = U_0 \cos(\omega t)$ while
the frequency is slowly swept through resonance. The signal
received from the fork is a current $I$ owing to the piezoelectric
effect. The stresses due to fork deflection induce charges and
thus the current is proportional to the derivative of the fork
deflection, \textit{i.e.} to the velocity:
\begin{equation}
I(t) = a\, \frac{dx(t)}{dt}~,
\label{adef}
\end{equation}
where $a$ is the fork constant. Its theoretical value is given
by\,\cite{Karrai2}
\begin{equation}
a =
3\,d_{11}\,E\,(\mathcal{T}\,\mathcal{W}/\mathcal{L})~,\label{theora}
\end{equation}
where $d_{11}\,=\,2.31\cdot10^{-12}$\,m/V is the longitudinal
piezoelectric modulus of quartz. For our forks $a =
2.47\cdot10^{-5}$\,C/m. The oscillation amplitude is therefore
known theoretically, but in practice the fork is usually
calibrated optically with interferometric
techniques.\cite{Karrai1,Karrai2,MagFields} Typically, only
\textit{ca.} 30\% of the theoretical current sensitivity is
achieved. In a cryogenic setup, where the fork is mounted inside a
sample container in the heart of the cryostat, a direct
measurement of the fork constant $a$ is complicated. Instead, we
prefer to calibrate our forks by comparing the mechanical
oscillator with the equivalent electrical RLC series resonance
circuit.\cite{MagFields} The corresponding differential equation
for the current is
\begin{equation}
\frac{d^2I}{dt^2}+\frac{R}{L}\frac{dI}{dt}+\frac{I}{LC}=\frac{1}{L}\frac{dU}{dt}~.\label{current}
\end{equation}
Comparing Eqs.~(\ref{current}) and (\ref{motion}) we see that
$\omega_0^{2}=1/(LC)$, $\gamma=R/L$, and using Eq.~(\ref{adef}),
$1/L = (F_0/U_0)\,a/m$. Additionally we have the condition that
the dissipated power at resonance has to be equal for both
equations: The electrical power $U_0^2/(2R)$ drives two legs of
the fork which dissipate $2\cdot F_0^2/(2 m \gamma)$. Thus we have
a closed set of equations which allows us to connect the
electrical and mechanical properties of the fork via the fork
constant $a$:
\begin{eqnarray}
F_0&=&(a/2)\,U_0~, \label{forceconv} \\
R &=& 2 m\gamma/a^2~, \label{r}\\
L &=& 2 m/a^2~,\\
C &=& a^2/(2k)~.
\end{eqnarray}
Experimentally the fork constant $a$ can be determined using
Eq.~(\ref{r}), which can be rewritten as
\begin{equation}
a = \sqrt{\frac{2m\,\Delta\omega}{R}}~. \label{expa}
\end{equation}
Here $\Delta\omega$ is determined from the width of the resonance
curve while $1/R$ is the linear slope of the experimental
$I_0(U_0)$ dependence, where $I_0$ is the current amplitude at
resonance. The only parameter which cannot be directly determined
from the experiment is the effective mass $m$. However the
theoretical value of the effective mass (Eq.~\ref{mvac}) seems to
be fairly reliable because of the close agreement between
theoretical and experimental values of the resonant frequency. The
example of our fork response in vacuum in Sec.~\ref{vacuum} leads
to $a=8.13\cdot10^{-6}\,$C/m, which amounts to 33\,\% of the
theoretical value.

In our experiment the absorption $I_{\rm a}(\omega)$ and dispersion $I_{\rm
  d}(\omega)$ components of the current $I(t) = I_{\rm a}
\cos(\omega t) + I_{\rm d} \sin(\omega t)$ are measured separately
with a lock-in amplifier (see Sec.~\ref{exper} for details). The
theoretical resonance curves,
\begin{eqnarray}
  I_{\rm a} &=& \frac{a^2 U_0}{2}\,\frac{m \gamma \omega^2}{(m \gamma \omega)^2
    + (m\omega^2 - k)^2} =
  \frac{I_0 (\Delta\omega)^2 \omega^2}{(\Delta\omega)^2 \omega^2
    + (\omega^2 - \omega_0^2)^2}
  ~, \label{Ia} \\
  I_{\rm d} &=& \frac{a^2 U_0}{2}\,\frac{\omega (m\omega^2 - k)}{(m \gamma \omega)^2
    + (m\omega^2 - k)^2} =
    \frac{I_0 \Delta\omega\, \omega\,(\omega^2 - \omega_0^2) }{(\Delta\omega)^2 \omega^2
      + (\omega^2 - \omega_0^2)^2} ~, \label{Id}
\end{eqnarray}
can be fit to the experimental response to determine the
parameters which enter Eq.~(\ref{motion}). In particular, the
absorption component $I_{\rm a}(\omega)$ reaches its maximum value
$I_0$ at a frequency which is exactly $\omega_0$ and the full
width $\Delta\omega$ of the absorption curve at 1/2 of the maximum
height $I_0$ gives exactly $\gamma$. If the fork constant $a$ is
known, then $m=a^2U_0/(2I_0\Delta\omega)$ and $k=m\omega_0^2$ can
be determined independently.

\section{INFLUENCE OF SURROUNDING MEDIUM ON THE OSCILLATING FORK}
\label{model}

\subsection{Hydrodynamic properties}\label{HydrodynProp}

In this section we outline the basic properties of the oscillatory
boundary layer flows, to understand how the vibrating fork works
as a detector. The classical viscous flow around a submerged
oscillating body\cite{LL} is rotational within a certain layer
adjacent to the body, while at larger distances it rapidly changes
to potential flow (if there is no free liquid surface or solid
surface in the vicinity of the oscillating body).  The depth of
penetration of the rotational flow is of order
\begin{equation}
\delta=\sqrt{\frac{2\nu}{\omega}}=\sqrt{\frac{2\eta}{\rho\omega}},
\end{equation}
where $\omega$ is the angular frequency of oscillation while
$\eta$ and $\nu=\eta/\rho$ are the dynamic and kinematic
viscosities of the fluid with density $\rho$.

As a result of the oscillatory motion of the body through the
liquid, the body experiences a force which has components
proportional to the velocity of the body $v$ (drag) and to its
acceleration $\dot v$ (mass enhancement):
\begin{equation}
{\cal F} = b v + \tilde m \dot v.
\label{hforce}
\end{equation}
To determine the values of $b$ and $\tilde m$ generally a full
solution of the flow field around the oscillating body is
required. Simplifications are possible in two limiting cases,
which depend on the relative magnitudes of the characteristic size
of the oscillating body $\ell$, oscillation amplitude $x_0$, and
viscous penetration depth $\delta$:

1) $\ell \ll \delta$ and $\omega x_0 \ell/\nu \ll 1$: In this case
the flow at any given instant can be regarded as steady -- as if
the body were moving uniformly with its instantaneous velocity. As
a rule, this case does not apply to the oscillating cryogenic
flows considered here.

2) $\ell \gg \delta$ and $\ell \gg x_0$: In this case the layer of
rotational flow around the body is very thin while in the rest of
the fluid the flow is potential. This case is directly applicable
to quartz tuning forks in $^3$He and $^4$He liquids: The kinematic
viscosity of normal liquid $^4$He-I above the $\lambda$-point is
$\nu_{4}\approx 2 \cdot 10^{-4}$~cm$^2$/s and of normal $^3$He
above the superfluid transition $\nu_{3}\approx 1$\,cm$^2$/s. At
$32$\,kHz we get the penetration depths $\delta_{4} \approx
0.4\,\mu$m and $\delta_{3} \approx 30\,\mu$m while $\ell \approx
400\,\mu$m for our forks. Moreover, the oscillation amplitude
would reach the leg thickness $\mathcal{T}$ only at a very high
velocity of order 1\,m/s ({\it cf.} Fig.~\ref{forkvac}). Note that
other oscillating objects such as spheres or wires may not always
be in this flow regime since they usually have smaller
characteristic size and smaller oscillation frequency.

When the conditions for case (2) above are valid, a major
contribution to both $b$ and $\tilde m$ in Eq.~(\ref{hforce}) is
found by solving the potential flow field $u$ around the
body.\cite{LL} In particular, $b$ is expressed as
\begin{equation}
b = \sqrt{\frac{\rho \eta \omega}{2}} \left[
\frac{1}{|v_0|^2}\oint|u_0|^2 dS \right] = \sqrt{\frac{\rho \eta
\omega}{2}}\,\mathcal{C}\,S~, \label{bcoeff}
\end{equation}
where $v_0$ and $u_0$ are the amplitudes of the velocities of the
body and the flow, the integral is taken over the surface of the
oscillating body, $S$ is the surface area of the body, and
$\mathcal{C}$ is some numerical constant which depends on the
exact geometry of the body. For example for a sphere
$\mathcal{C}=3/2$, while for an infinitely long cylinder
oscillating perpendicular to its axis $\mathcal{C}=2$.

The largest contribution to mass enhancement $\tilde m$ comes from
the potential flow around the body and can be expressed through
the mass $\rho V$ of the liquid displaced by the body of volume
$V$. A smaller contribution is caused by the fact that the viscous
drag force experienced by the body is usually phase shifted with
respect to the velocity of the body. This can be interpreted such
that a volume of order $S \delta$ of the liquid is clamped to
comotion with the oscillating body. Thus for $\tilde m$ we can
write
\begin{equation}
\tilde m = \beta \rho V + B \rho S \delta~,
\label{hydrom}
\end{equation}
where $\beta$ and $B$ are again geometry-dependent coefficients.
For example, for a sphere\cite{LL} $\beta = 1/2$ and $B = 3/4$;
for an infinitely long cylinder with elliptic cross
section\cite{MT} $\beta = r_\perp/r_\parallel$, where $r_\perp$
and $r_\parallel$ are the lengths of the axes which are
perpendicular and parallel to the oscillation direction,
respectively; for a rectangular beam\cite{sader} $\beta = (\pi/4)
a_\perp/a_\parallel$, where $a_\perp$ and $a_\parallel$ are the
lengths of the sides which are perpendicular and parallel to the
oscillation direction, respectively.

We are not aware of rigorous calculations of the parameters
$\beta$, $B$, and $\mathcal{C}$ for a tuning fork. A single
oscillating beam has been considered before in great
detail.\cite{sader} However, the presence of two legs in close
vicinity of each other significantly affects the potential flow
field and changes, for example, the $\beta$ parameter.\cite{Clubb}
Thus we consider $\beta$, $B$, and $\mathcal{C}$ as fitting
parameters, to be determined for a particular fork from the
experiment. This approach was previously used, for example, in
Ref.~\onlinecite{martinez} for a vibrating reed in liquid $^4$He
and it seems to be provide the first step for understanding the
experimental results. \vspace{-5mm}

\subsection{Hydrodynamic model of sensor}

The addition of the force $\cal F$ to the equation of motion of
the fork (\ref{motion}) leads to a reduction in the resonant
frequency and an increase in the width of the resonance curve:
\begin{eqnarray}
\omega_0^2 &=& \omega_{0\rm vac}^2\,(m_{\rm vac}/m),\\
\gamma &=& \gamma_{\rm vac}(m_{\rm vac}/m) + b/m,
\end{eqnarray}
where $m = m_{\rm vac} + \tilde m$ is the effective mass of the
oscillating body immersed in the fluid. For convenience we
redefine the fork parameters $\beta$ and $B$ from
Eq.~(\ref{hydrom}) as relative to the effective fork mass in
vacuum:
\begin{equation}
\tilde m = m_{\rm vac} \; \frac{\rho}{\rho_{\rm q}} \left[\beta +
B \delta (S/V)\right]~.
\end{equation}
Ignoring the vacuum resonance width $\Delta f_{\rm vac}$, we
finally obtain the dependence of the resonant frequency $f_0$ and
of the full width of the absorption curve at half height $\Delta
f$ on the fluid density and viscosity:
\begin{eqnarray}
\left(\frac{f_{0\rm vac}}{f_0}\right)^2 &=& 1+
\frac{\rho}{\rho_{\rm q}} \left(\beta + B\, \frac SV\,
\sqrt{\frac{\eta}{\pi \rho f_0}}
\right),\label{f0}\\[1ex]
\Delta f &=& \frac12 \sqrt{\frac{\rho \eta f_0}{\pi}} \,
\mathcal{C}\,S\,\frac{(f_0/f_{0\rm vac})^2}{m_{\rm
vac}}.\label{df}
\end{eqnarray}
Here $V=\mathcal{TWL}$ and $S=2(\mathcal{T + W)L}$.

Eqs.~(\ref{f0}) and (\ref{df}) can now be used to determine
experimentally the hydrodynamic parameters from measurements in a
fluid with known $\rho$ and $\eta$. Once the parameters are known,
the fork can be used for measurements of $\rho$ and $\eta$, in
principle, for any other medium and thus as a pressure and
temperature sensor, if $\rho(P,T)$ and $\eta(P,T)$ are known. From
this point of view, the width $\Delta f$ is especially useful for
measurements as it requires calibration of only one parameter
($\mathcal{C}/m_{\rm vac}$). In Eq.~(\ref{f0}) the factor
multiplying the parameter $B$ is small and so, even here, often
only one parameter $(\beta)$ is of major importance.
Unfortunately, our measurements indicate that the parameters vary
from one fork to the next and calibrations need to be checked (see
Secs.~6 and 7). The calibration should be re-examined even when
performing measurements with the same fork in widely differing
conditions (see Sec.~6). \vspace{-5mm}

\subsection{Beyond the model of viscous hydrodynamics}

With decreasing temperature the mean free path of excitations
increases in both $^4$He and $^3$He superfluids and the
hydrodynamic description ceases to be valid as the normal fluid
penetration depth grows beyond all relevant length scales. In
$^4$He-II the crossover to the ballistic regime takes place below
1\,K and in $^3$He-B below $0.3\,T_{\rm c}$. At low temperatures
the drag is caused by the scattering of the excitations from the
oscillating body. As the excitation density decreases with
decreasing temperature the drag coefficient also rapidly
decreases: $b \propto T^4$ for phonons in $^4$He-II, while $b
\propto \exp(-\Delta/k_{\rm B} T)$ for rotons and for
quasiparticles in $^3$He-B, where $\Delta$ is the relevant energy
gap. The crossover from the hydrodynamic to the ballistic regime
has been described for a vibrating wire in
Refs.~[\onlinecite{Morishita,BunkovVWR}] and for a vibrating
sphere in [\onlinecite{WSPhysica}]. In this work we are not
discussing the ballistic regime further.

In everyday life a tuning fork is used to create sound in air. In
the case of a quartz tuning fork in LHe we might wonder whether
the compressibility and the losses from sound emission need to be
taken into account. A quartz fork operates at higher frequency
than vibrating wires, grids, spheres and most other oscillating
bodies. Therefore sound emission might be sizable for the fork
while it is negligible in these other cases. The power loss from
acoustic emission reduces the $Q$ value and contributes to the
width of the resonance curve $\Delta\omega_{\rm ac} = R_{\rm
ac}/m$, where the average power loss ${1\over 2} R_{\rm ac} \,
v_0^2$ has been expressed in terms of the so-called radiation
resistance $R_{\rm ac}$.

A realistic calculation of the acoustic emission from a tuning
fork is complicated. Clubb \textit{et al.}\cite{Clubb} suggest a
model of two infinite cylinders oscillating at 180$^\circ$
out-of-phase and give for their quadrupolar acoustic field the
radiation resistance
\begin{equation}
R_{\rm ac} = \frac{\pi^2 \omega^5 \rho\, \mathcal{W}^6
\mathcal{L}}{11616 \, c^4}~. \label{Rac}
\end{equation}
This expression gives a very small contribution to the resonance
width and, owing to the high powers in which the different
quantities appear, quantitative comparison with experiment is so
far inconclusive. In our measurements on normal $^3$He a small
temperature independent constant contribution to the resonance
width is distinguishable at high temperatures (see Sec.6).
However, its magnitude is not in agreement with Eq.~(\ref{Rac})
and at this point the presence of the acoustic loss term remains
unclear. \vspace{-3mm}

\section{SENSOR PREPARATION AND MEASUREMENT} \label{exper}

\begin{figure}[t]
\centerline{\includegraphics[width=0.7\linewidth]{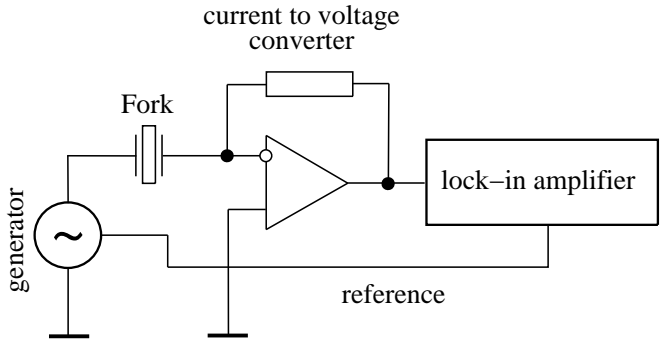}}
\caption{Circuit diagram for tuning fork measurements. }
\label{scheme}
\end{figure}

A commercial quartz tuning fork comes in a vacuum-tight sealed
metal can which has to be partly or entirely removed, to probe the
flow properties of the surrounding medium. The pair of leads for
exciting and sensing the oscillator is magnetic. In magnetically
sensitive applications they have to be removed and changed to
nonmagnetic leads. For measuring the current in
Eq.~(\ref{current}) one usually uses the current input of a
phase-sensitive lock-in amplifier. The excitation voltage is
supplied by a high resolution digital generator, which also
provides the reference signal for the lock-in amplifier (see
circuit diagram in Fig.~\ref{scheme}).

\begin{figure}[t]
\centerline{\includegraphics[width=0.75\linewidth]{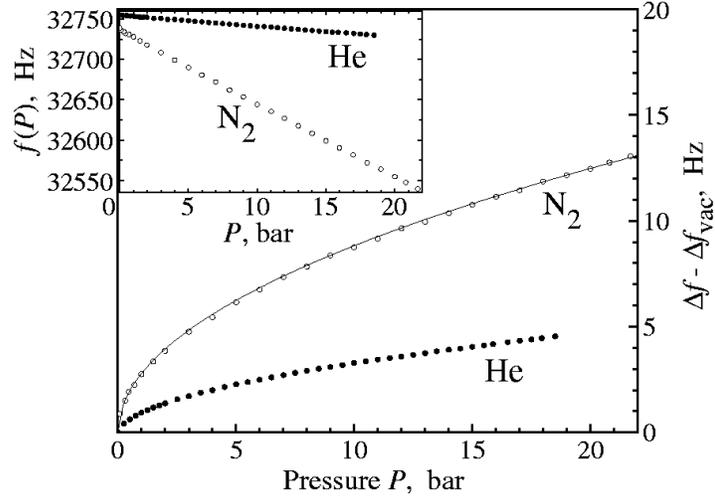}}
\caption{Room temperature tests of sensor sensitivity.
\textit{(Main panel)} Resonance width $\Delta f$, with the vacuum
width $\Delta f_{\rm vac}$ subtracted (right vertical scale),
plotted versus applied pressure in gaseous nitrogen
($T=22.5^\circ$C) and helium ($T=23^\circ$C). The solid line is
the fitted square root dependence, as expected from
Eq.~(\protect\ref{df}). \textit{(Inset)} The corresponding
resonant frequencies $f_0$ are almost linear with pressure, as
expected from Eq.~(\protect\ref{f0}). The data for N$_2$ and He
have been measured with two different forks. } \label{room}
\end{figure}

In Fig.~\ref{room} the results are plotted from room temperature
test measurements, to compare to Eqs.~(\ref{f0}) and (\ref{df}).
The measured nearly linear dependence of the resonant frequency
and the square root dependence of the resonance width versus
applied pressure $P$ follow directly from these equations,
assuming that $\rho \propto P$ and $\eta$ does not appreciably
change with $P$. In these measurements we also tested if the
results showed variations depending on whether (i) only a small
hole is made in the encapsulating can, or (ii) the entire top
surface of the can is ground away, or (iii) the can is completely
removed. No obvious qualitative differences were observed, which
is as expected, since for all data in Fig.~\ref{room} the
penetration depth $\delta$ is much smaller than the fork
dimensions.

\begin{figure}[t]
\centerline{\includegraphics[width=0.67\linewidth]{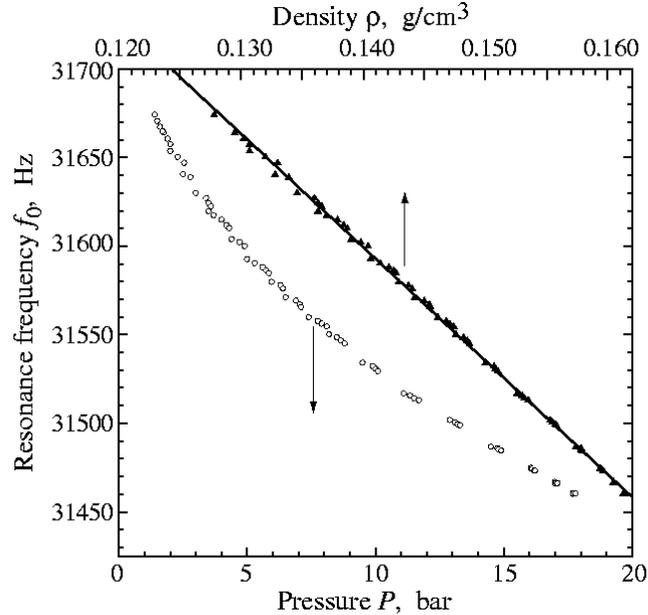}}
\caption{Resonant frequency $f_0$ in liquid $^4$He at 4.2\,K
plotted versus applied pressure (bottom) and density (top). The
tuning fork was in its original can, but with the flat top surface
of the can ground away and $\Delta f_{\rm vac} = 0.06\,$Hz. The
pressure was converted to density using the HEPAK
package.\cite{HEPAK} The solid line is a linear fit through the
data. } \label{liquidHePressure}
\end{figure}

In Fig.~\ref{liquidHePressure} a similar measurement has been
performed in liquid $^4$He at 4.2\,K as a function of pressure.
Owing to the non-linear dependence of the density of liquid He on
applied pressure the resonant frequency shows linear dependence
only after converting pressures to densities. For these results
thermal aging is of some concern. We tested the stability of the
resonant frequency and width of five different forks by cycling
them between 300 and 77\,K, with their cans completely removed.
For a virgin sensor resonant frequency shifts of $\lesssim
0.2\,$Hz and changes in width $\lesssim 0.3\,$Hz are typical. For
most applications such shifts are negligible, but if more stable
reproducible results are required, then the sensor should be
thermally cycled. After a few cycles the changes are considerably
reduced, but only after 30 -- 50 cycles the results become stable.
It should be noted that larger changes can result from bending the
leads of the fork or from making new solderings to the leads,
presumably because new strains are imposed via the electrodes on
the quartz surface. Such changes, which can be $\lesssim 0.7\,$Hz
in both resonant frequency and width, are generally larger than
those caused by the removal of the can by grinding. This suggests
that a fork, which is used as thermometer, should be handled with
care.

A further practical consideration in connection with
Fig.~\ref{liquidHePressure} is that these results could only be
obtained after the tuning fork was installed in an isolated sample
container in controlled conditions. In a LHe bath in an open dewar
surface conditions on the fork may change because of adsorbed gas
or particles floating in the bath after a LHe transfer. Typically
in such conditions the resonance width does not stay constant, but
gradually increases during a long run. Occasionally small step
changes in the resonant frequency are observed which could arise
if air flakes stick on the fork, for instance. Thus for accurate
and reproducible readings the fork should preferably not be used
in technical helium.

\begin{figure}[t]
\begin{center}
\includegraphics[width=100mm]{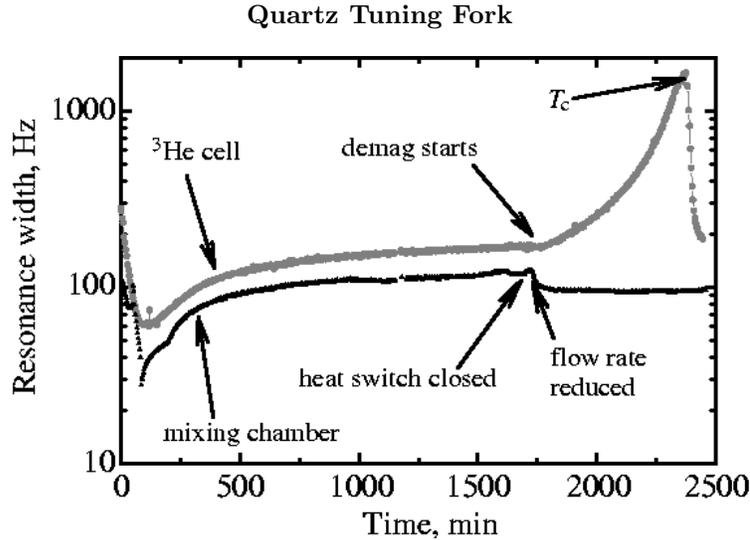}
\caption{Two quartz tuning forks monitoring temperatures in a
nuclear cooling cryostat. The bottom trace shows the sensor inside
the mixing chamber of the dilution refrigerator in the
concentrated $^3$He--$^4$He solution. The top trace gives the
simultaneously measured width of the sensor on the nuclear cooling
stage inside the liquid $^3$He sample container at zero liquid
pressure. } \label{precool}
\end{center}
\end{figure}

Fig.~\ref{precool} shows two examples of the quartz tuning fork as
a practical monitoring device of temperatures in a nuclear cooling
cryostat.\cite{Pobell,stage} The two traces illustrate the range
which the resonance width $\Delta f$ traverses when the cryostat
is taken through its cooling cycle. The bottom trace represents a
sensor in the mixing chamber, while the top trace monitors one on
the nuclear cooling stage. The most prominent features in these
traces are abrupt anomalies: In the bottom trace from
disconnecting the superconducting heat switch between the mixing
chamber and the nuclear cooling stage. Here the mixing chamber
first starts cooling, but then warms up to a new higher level when
the $^3$He flow rate in the dilution refrigerator circulation is
reduced. Simultaneously the demagnetization of the nuclear cooling
stage is started. This means that in the upper trace the
temperature starts decreasing and $\Delta f$ increasing. When the
cool down passes through the superfluid transition temperature
$T_{\rm c}$, the width suddenly dives in a rapid decrease (see
Sec.~6). These recordings show that the quartz tuning fork
provides rapid and sensitive confirmation of the actions which are
performed on the cryostat.

\section{PROPERTIES OF TUNING FORKS IN VACUUM} \label{vacuum}

\begin{figure}[t]
\centerline{\includegraphics[width=\linewidth]{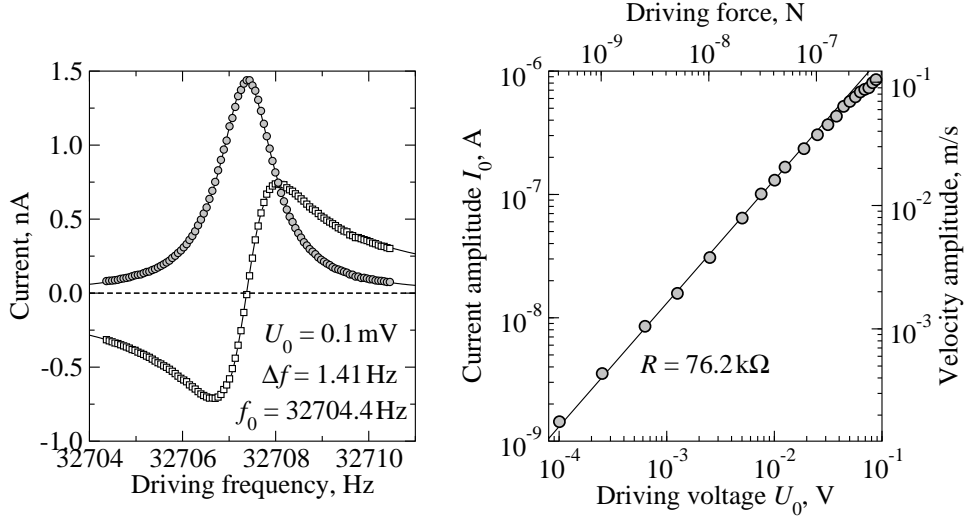}}
\caption{Response of the quartz tuning fork in vacuum at a
temperature of about 1\,K. \textit{(Left)} Resonance curves for
absorption $I_{\rm a}$ ($\circ$) and dispersion $I_{\rm d}$
($\square$) components of the current measured at fixed excitation
level $U_0$. The solid curves are fits to Eqs.~(\ref{Ia}) and
(\ref{Id}). Background contributions from a shunting capacitance
and leads are subtracted. \textit{(Right)} Dependence of the
amplitude $I_0$ of the current at resonance on the excitation
voltage amplitude $U_0$. Conversions to velocity and driving force
via Eqs.~(\ref{adef}) and (\ref{forceconv}) are also shown.}
\label{forkvac}
\end{figure}

In its original package the quartz tuning fork comes inside a
vacuum-tight can. When the fork is cooled in this can, the
resonant frequency decreases and the $Q$ value increases. At LHe
temperatures the reduction in resonant frequency from the
room-temperature value is about 70\,Hz and the $Q$ value
approaches $10^6$. If the can is removed the bare fork behaves in
vacuum in similar manner: The reduction in the resonant frequency
remains the same while the $Q$ value is typically lower than for
the fork inside its original can. However, the $Q$ value varies
greatly from one fork to the next. Typical $Q$ values range from
$2\cdot 10^4$ to $5\cdot 10^5$. A similar large scatter was
observed in previous reports.\cite{MagFields,Clubb} The reasons
for such variability between different forks are not known. In
principle differences may result from slight damage (when the can
is removed, for instance) or from dirt accumulated on the legs. In
practice of course, the resonance width is much larger in LHe than
the vacuum width and thus these problems do not affect our later
analysis. However, if one is interested in the properties at the
lowest temperatures in the ballistic regime, then special care has
to be invested in selecting forks with the narrowest possible
vacuum width.

The vacuum response of a tuning fork at $\sim 1\,$K is shown in
Fig.~\ref{forkvac}. This fork was later used for measurements in
$^3$He liquid at zero pressure. The data in Fig.~\ref{forkvac}
were measured at different excitation levels in the $^3$He sample
container while the $^3$He pressure was less than 1\,mbar at a
temperature of around 1\,K. As mentioned above (Sec.~2.2), from
the resonance characteristics measured at various excitation
voltage amplitudes $U_0$ it is possible to determine the fork
constant $a$ using Eq.~(\ref{expa}). In this case $a = 8.13\cdot
10^{-6}\,$C/m. Now the driving force $F_0$ and the velocity
amplitude can be calculated from Eqs.~(\ref{forceconv}) and
(\ref{adef}), respectively.  Thus the measurement of $I_0$ versus
$U_0$ in the right panel of Fig. \ref{forkvac} can be converted to
a dependence of the tuning fork velocity amplitude on the driving
force.

As seen in Fig.~\ref{forkvac}, in vacuum the fork responds
linearly up to a driving force of order 100\,nN, which corresponds
to displacements of a few $\mu$m. Above this limit the amplitude
in the motion of the legs starts to be sufficient for the response
to become nonlinear. The cause for the nonlinear behavior is
deformation in the sensor material during large-amplitude
oscillation. The oscillation amplitudes of ions around the minimum
of their potential become large enough such that anharmonic terms
in the potential energy introduce nonlinear restoring forces in
the ion motion. The overall effect is the appearance of a
nonlinear restoring force in the equation of motion,
Eq.~(\ref{motion}). In addition at low temperatures other sources
contribute to nonlinearities in large amplitude oscillation, such
as slow strain release from defects in the oscillator material.
These have been extensively investigated, for instance with
vibrating wire resonators in vacuum.\cite{pobell} The nonlinear
drive regime of the quartz fork is not discussed in this report.

\section{TUNING FORK IN $^3$He} \label{he3}

Our measurements in liquid $^3$He have been performed in two
different cryostats and with different sensors.\cite{forks-he3} In
the measurements at 29\,bar pressure, the fork parameters were
$f_{0\rm vac} = 32705.05\,$Hz and $\Delta f_{\rm vac} = 0.06\,$Hz
in vacuum at LHe temperatures. In this setup temperatures above
$T_{\rm c}$ are measured by a melting curve thermometer which is
mounted on the nuclear cooling stage. Below $T_{\rm c}$, the
temperature readings are determined from nuclear magnetic
resonance frequency shifts of the $^3$He sample.\cite{ahonen} The
NMR reading is preferred below $T_{\rm c}$ because it is measured
directly from the liquid in which the tuning fork is also
immersed. This minimizes thermal gradients between the thermometer
and the fork.

In the zero pressure measurements the fork had $f_{0\rm vac} = 32707.4\,$Hz
and $\Delta f_{\rm vac} = 1.41\,$Hz. In this setup the temperature is
determined with pulsed NMR on Pt powder immersed in the liquid $^3$He
sample. The NMR signal amplitude is calibrated using the known value of the
superfluid $^3$He transition temperature $T_{\rm c}$. The superfluid
transition is indicated by the fork reading or by two additional vibrating
wire resonators in the $^3$He cell which also are used for thermometry.

\begin{figure}[t]
\centerline{\includegraphics[width=0.9\linewidth]{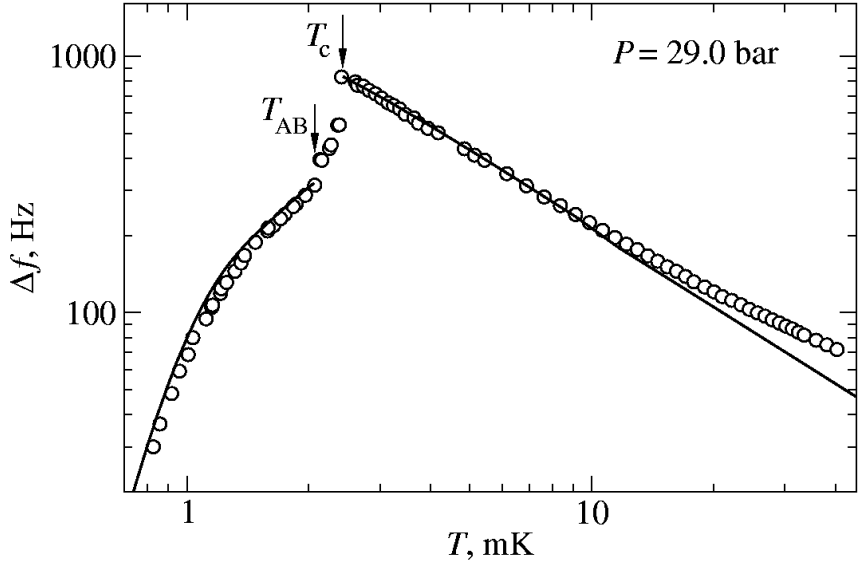}}
\caption{Resonance width of the quartz tuning fork in liquid
$^3$He at 29 bar pressure. The measured data are marked with
circles. A sharp reduction in the width is observed on cooling
below $T_{\rm c}$ and an abrupt discontinuity at the AB
transition. The solid line is the predicted
behavior\cite{BunkovVWR} of a vibrating wire resonator with the
same density $\rho_{\rm q}$ and the same vacuum resonant frequency
$f_{0\rm vac}$ as the fork. To produce a good fit to the quartz
fork data the wire diameter had to be fixed to 0.25\,mm which is
comparable to the dimensions of the legs of the fork.}
\label{he3allT}
\end{figure}

The temperature dependence of the resonance width $\Delta f$ at
29\,bar pressure is presented in Fig.~\ref{he3allT}. The
measurements were performed in the linear drive regime of the fork
with the maximum current not exceeding 4\,nA (and thus velocities
not exceeding 0.5\,mm/s).  The plot demonstrates that the width
changes rapidly in the range 0.8\,--\,40\,mK and, once calibrated,
can be used as a thermometer. The measurements were extended to
much lower temperatures than shown in Fig.~\ref{he3allT}, the
lowest observed resonance width  was $\Delta f = 0.5\,$Hz.
However, at these temperatures we have no other thermometer in the
$^3$He sample container to calibrate the fork.

Fig.~\ref{he3allT} can be divided in three temperature regimes:
(i) normal $^3$He above $T_{\rm c}$, where the width rapidly
increases with decreasing temperature due to the Fermi-liquid
behavior of the viscosity, $\eta \propto 1/T^2$, (ii) superfluid
$^3$He-A and (iii) $^3$He-B phases, where the width decreases with
decreasing temperature mainly because of the decreasing
normal-fluid density. Fig.~\ref{he3allT} provides an interesting
comparison of the fork oscillator with the vibrating wire
resonator. In Ref.~[\onlinecite{BunkovVWR}] the available
theoretical and experimental information on vibrating wires has
been combined in a computer program which calculates the response
of a vibrating wire loop in normal $^3$He and in superfluid
$^3$He-B for a resonator with known wire diameter, wire density,
and resonant frequency in vacuum. In Fig.~\ref{he3allT} the
response of a fictitious wire loop with the density of quartz and
the frequency of the tuning fork in vacuum has been fitted to the
experimental data with the wire diameter as a fitting parameter.
The fit is remarkably good and gives a reasonable value for the
wire diameter which is of order of the thickness of the fork leg.
The same fit does not reproduce exactly the temperature dependence
of the resonant frequency owing to the difference in the $\beta$
factor between the wire and the fork (\textit{cf.} Eqs.~(\ref{f0})
and (\ref{df})). The comparison is  used here to emphasize that
the fork thermometer is very comparable to the vibrating wire in
this temperature range of liquid $^3$He.

\begin{figure}[t]
\centerline{\includegraphics[width=0.9\linewidth]{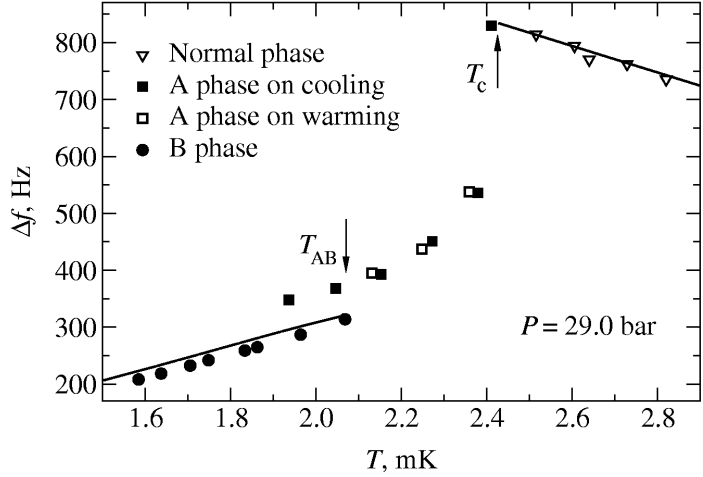}}
\caption{Closeup of resonance width in Fig.~\protect\ref{he3allT}
in the temperature region of $^3$He-A. During cooling the
A$\rightarrow$B transition is supercooled, while during warming
the B$\rightarrow$A transition occurs close to the thermodynamic
equilibrium temperature $T_{\rm AB}$ of this first order phase
transition. The discontinuous jump in resonance width at the
transition is mainly caused by the change in $\rho_{\rm n}$. The
solid lines depict the vibrating wire model of
Fig.~\protect\ref{he3allT}.} \label{he3close-up}
\end{figure}

Vibrating wires are not generally used as thermometers in $^3$He-A
owing to their texture-dependent non-reproducible response. Our
limited experience with forks shows that for a given fork the
width in the A phase is reproducible, independently whether one
enters the A phase from the normal phase or from the B phase
direction, as seen in Fig.~\ref{he3close-up}. Probably the larger
dimensions of the fork legs fix the orientations in the order
parameter texture such that the response becomes reproducible if
the oscillation amplitude remains small compared to the leg
dimensions. Nevertheless, more measurements are required to
establish whether tuning forks can be used as accurate secondary
thermometers also in $^3$He-A.

\begin{figure}[t]
\centerline{
\includegraphics[height=6cm]{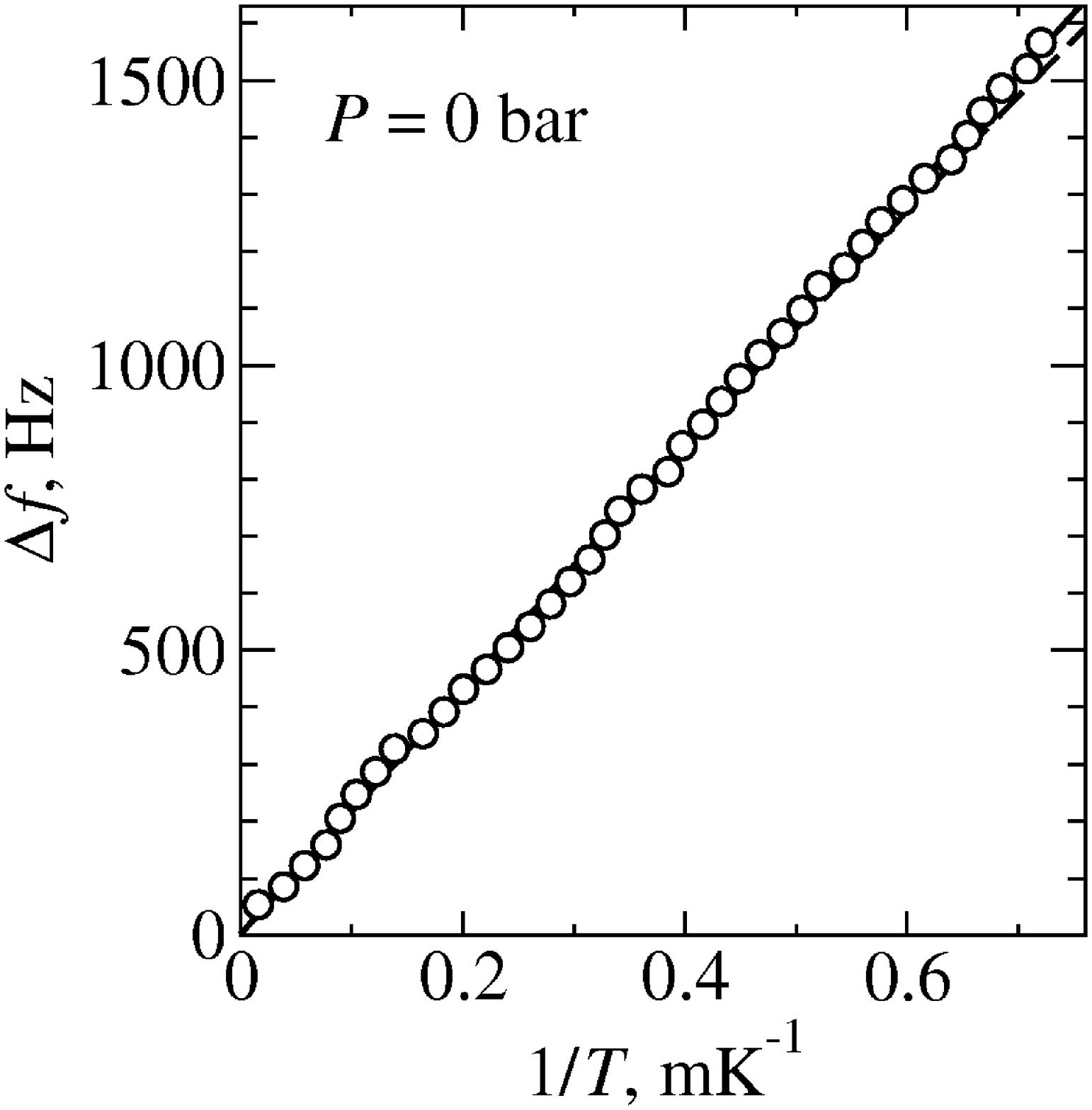}
\hfill
\includegraphics[height=6cm]{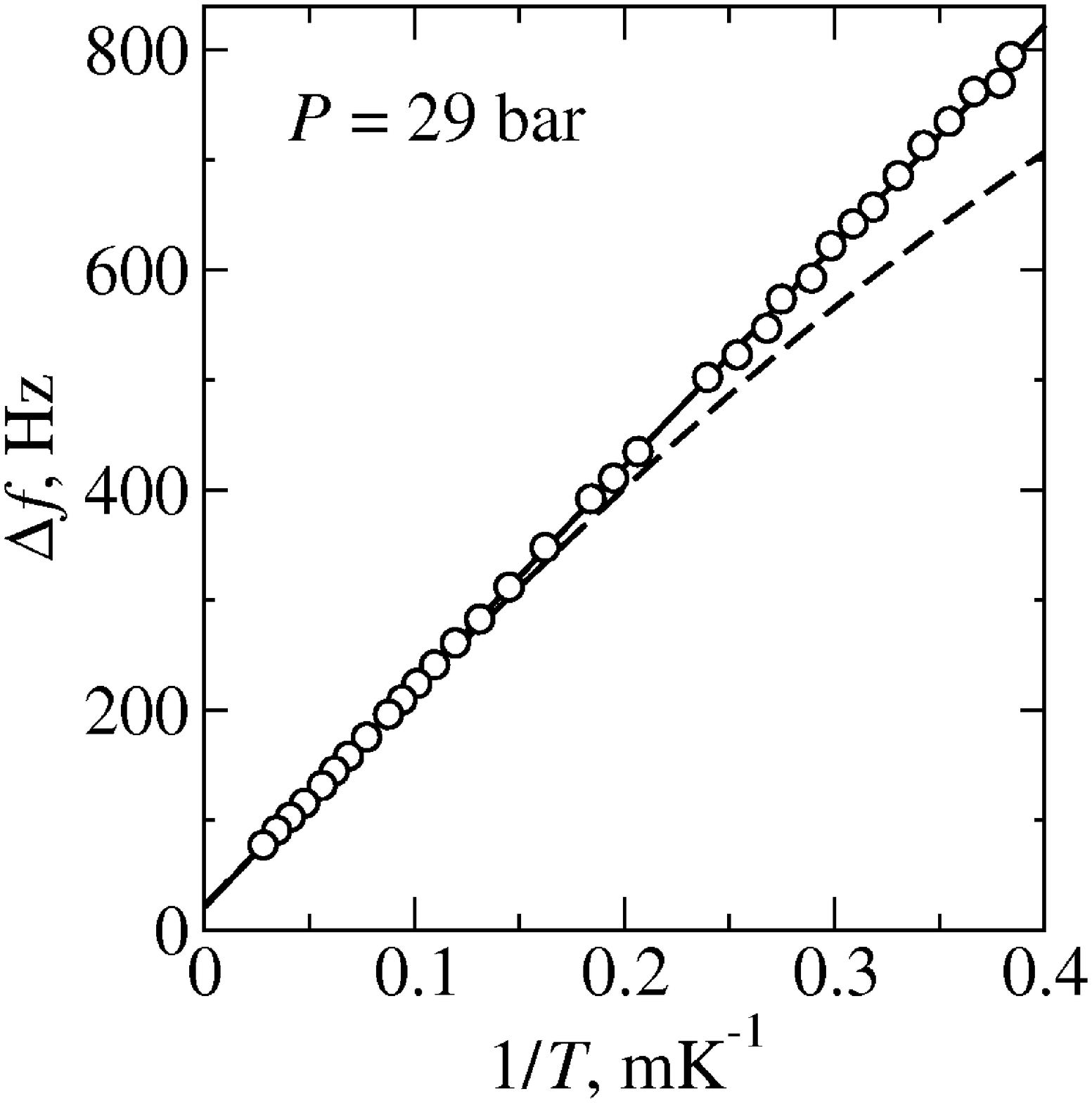}
} \caption{Resonance width at two different pressures in normal
$^3$He as a function of inverse temperature. The experimental data
is shown as circles. The solid lines are linear fits of $\Delta f$
versus $1/T$. For comparison, the dashed curves show the effect
from a viscosity anomaly close to $T_{\rm c}$ which was measured
with a vibrating wire resonator in Ref.~[\onlinecite{carless}].}
\label{he3norm}
\end{figure}

In Fig.~\ref{he3norm} the fork properties are analyzed in more
detail in normal $^3$He. Since the viscosity of normal $^3$He
varies as $\eta \propto T^{-2}$, we expect the width to depend on
temperature as $\Delta f \propto T^{-1}$, Eq.~(\ref{df}). This
dependence is indeed observed in the experiment. However, a
constant addition to the resonance width is also present in the
experimental data (seen as a non-zero intercept on the vertical
scale in the two panels of Fig.~\ref{he3norm}). This additional
temperature-independent contribution to the width is 21.2\,Hz at
29\,bar and 3.2\,Hz at zero pressure. In Fig.~\ref{he3allT}, this
effect appears  as a tendency towards a constant width at high
temperatures.

The origin of the temperature-independent contribution to the
width in normal $^3$He is not clear. Clubb et al.\cite{Clubb}
observed the same effect in $^3$He--$^4$He mixtures and attributed
it to acoustic emission. Their model, Eq.~(\ref{Rac}), gives  an
orders of magnitude smaller value than the measured one in our
case. Moreover, irrespectively of the model one would expect that
the losses from sound emission will be smaller at high pressures
(since the sound velocity increases with pressure), while the two
measurements in Fig.~\ref{he3norm} show opposite behavior. Since
these have been performed with different sensors, the possibility
remains at this point that the main part of the temperature
independent width in normal liquid $^3$He depends on the fork with
which it is measured.

Several reports on vibrating wire measurements mention a viscosity
anomaly in normal $^3$He: an unexpected reduction in the viscosity
close to $T_{\rm c}$ from the $T^{-2}$ behavior.\cite{Dobbs} Our
data show no sign of this anomaly: The width $\Delta f$ changes
exactly proportional to $T^{-1}$ until $T_{\rm c}$
(Fig.~\ref{he3norm}). The reason for this difference is not clear.
Possibly the fork owing to larger size and higher frequency
operates in a different hydrodynamic regime than typical vibrating
wire resonators. In particular close to $T_{\rm c}$, where the
viscosity of $^3$He is the highest, the viscous penetration depth
becomes comparable to the characteristic size of the oscillating
object, especially at low pressure. For example for a fork at
$P=0$ and $T=1$\,mK, $\delta$ is about the inter-leg distance
$\cal D$. Thus an interpretation of the results in terms of the
simple model presented in Sec.~3 may not be justified. The
question which kind of viscometer is more appropriate for $^3$He
at temperatures close to $T_{\rm c}$ requires further analysis.

From the linear fit in Fig.~\ref{he3norm} the fork parameter
$\mathcal{C}$ in Eq.~(\ref{df}) can be determined. Using viscosity
data from Ref.~[\onlinecite{carless}] (omitting the viscosity
anomaly close to $T_{\rm c}$) we get $\mathcal{C} = 0.57$ for the
fork used at zero pressure and $\mathcal{C} = 0.64$ for the fork
used at 29\,bar .

\begin{figure}
\centerline{\includegraphics[width=0.7\linewidth]{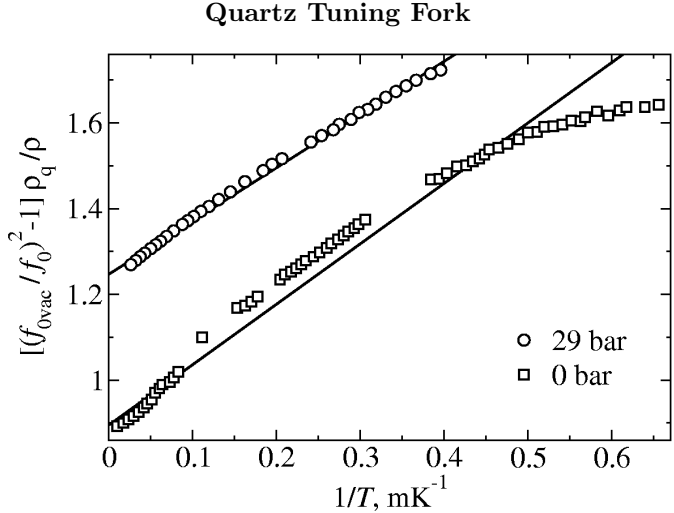}}
\caption{Resonance frequencies of the two quartz tuning forks from
Fig.~\protect\ref{he3norm} in normal $^3$He, plotted as $[(f_{0
\rm{vac}}/f_0)^2 - 1]\rho_{\rm q}/\rho$ versus $1/T$. The symbols
represent experimental data. The solid lines are fits to
Eq.~(\ref{f0}), assuming $\eta \propto T^{-2}$.} \label{he3freq}
\vspace{-5mm}
\end{figure}

The relative change in the density of liquid $^3$He between 0 and
40\,mK is about $10^{-4}$, according to Ref.~[\onlinecite{Roach}].
Thus the largest contribution to the temperature dependence of the
resonant frequency of the fork in this temperature range comes
from the viscous mass enhancement in Eq.~(\ref{f0}). Our
experimental data for normal $^3$He together with a fit to
Eq.~(\ref{f0}) are shown in Fig.~\ref{he3freq}. The quantity
plotted on the vertical scale is the term in the parentheses on
the right hand side of Eq.~(\ref{f0}). The fit gives $\beta =
0.90$ and $B = 0.73$ for the mass enhancement parameters of the
fork used at zero pressure and $\beta = 1.25$ and $B = 1.05$ for
the fork used at 29 bar. The $\beta$ factor can also be determined
from measurements at the very lowest temperatures in $^3$He-B,
well in the ballistic regime, when the shift of the resonant
frequency is caused entirely by inertial effects. This way we
obtain $\beta = 0.88$ for the fork used at zero pressure and
$\beta = 1.20$ for the fork used at 29 bar. These second values
are close to the ones shown as the zero intercepts on the vertical
scale of Fig.~\ref{he3freq}.

We conclude that the mass enhancement factors $\beta$ and $B$ turn
out to have rather different values for our two forks, in spite of
the fact that these two forks have closely similar dimensions and
room temperature oscillator properties. In both cases $\beta$ is
larger than the theoretical value $\beta = (\pi/4) \mathcal{W/T} =
0.69$ for a single beam of the size of one leg, as might be
expected owing to the two legs of the fork in close vicinity of
each other, Sec.~\ref{HydrodynProp} The sizeable differences in
the fitted parameter values do not support our hopes that quartz
tuning forks could be used as reproducible secondary thermometers,
with a common calibration for all forks of the same type.

Another important question is whether for a given fork the
calibration obtained for one pressure can be used to interpret
results at another pressure without re-calibration. To check this
we repeated measurements at zero pressure using the fork for which
a calibration at $P=29$\,bar had been obtained. In the regime
$\delta \ll ({\cal D}, {\cal T}, {\cal W})$, where the simple
model from Sec.~3.1 is applicable and which at zero pressure
corresponds to $T \gtrsim 3\, T_{\rm c}$, the measured resonance
frequency and width are within 10\% from the prediction of the
model. When temperatures approach $T_{\rm c}$ the deviation
increases and at $T_{\rm c}$ the resonance width is twice larger
than expected. In this temperature range evidently the interaction
between the two legs of the fork becomes important and probably
the surrounding of the fork (which here includes another fork less
than 1\,mm away) also influences the result. Thus we can conclude
that the simple hydrodynamic model presented here describes
reasonably well the behavior of the fork in normal $^3$He but is
not sufficient for exact scaling of a temperature calibration from
one pressure to another, especially in the regime of large viscous
penetration depth close to $T_{\rm c}$. \vspace{-5mm}

\begin{figure}[t]
\centerline{\includegraphics[width=0.73\linewidth]{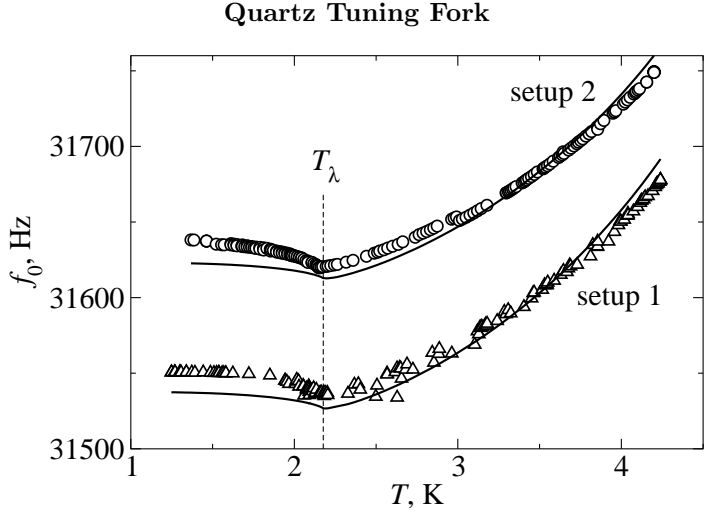}}
\caption{Resonance frequencies of two quartz tuning forks in
liquid $^4$He at saturated vapor pressure, measured in two
different setups. The symbols represent experimental data. The
lines have been fit to the data at $T>T_\lambda$, using
Eq.~(\ref{f0}) with $\beta$ as fitting parameter and $B=1$. The
fit gives $\beta = 1.39$ for setup 1 and $\beta = 1.27$ for setup
2. The data on the physical properties of liquid $^4$He are from
Ref.~[\onlinecite{Russ}].} \label{he4freq}
\end{figure}

\section{TUNING FORK IN $^4$He} \label{he4}

Our measurements in liquid $^4$He have also been performed in two
different setups. In both cases the forks\cite{forks-he4} are
partially inside their original cans, only a hole is ground in the
can to provide a connection between the fork and the liquid in the
$^4$He sample container. The temperature is determined from the
saturated vapor pressure. In the first setup the LHe temperature
vacuum parameters of the fork are $f_{0\rm vac} = 32708\,$Hz and
$\Delta f_{\rm vac} = 0.4\,$Hz.  In the second setup a fork is
used with $f_{0\rm vac} = 32709.97\,$Hz and $\Delta f_{\rm vac} =
0.06\,$Hz.

The resonance frequencies of the two forks are shown in
Fig.~\ref{he4freq} as a function of temperature. In liquid $^4$He
above the superfluid transition the density changes faster than
the viscosity. Thus the resonant frequency is more useful for
thermometry. Indeed the measured resonant frequency is reminiscent
of the inverse of the well-known liquid density, with a maximum in
the density just above the superfluid transition $T_{\lambda}$.
However, a fit of the measured frequency in the normal phase to
Eq.~(\ref{f0}) shows systematic differences which so far remain
unexplained. In normal $^4$He the viscous penetration depth
$\delta$ is of sub-micron size. Thus the influence of the small
viscous term in the added mass cannot be reliably distinguished in
the presence of the rapid variation of the liquid density.
Therefore, in Fig.~\ref{he4freq} the value of $B$ is fixed to 1
and the only fitting parameter is $\beta$.

Below $T_\lambda$ we plot the extrapolation of Eq.~(\ref{f0}) in
Fig.~\ref{he4freq}, assuming a simple two-fluid-model
interpretation. The interaction of the normal and superfluid
fractions via the possible existence of quantized vortices is
neglected. The inertial contribution to the effective mass is
attributed to the whole fluid while the viscous contribution is
assumed to originate only from the normal fluid component. Thus
$(f_{0\rm vac}/f_0)^2 = 1+ \beta\rho/\rho_{\rm q} + B S/(V
\rho_{\rm q}) \sqrt{\eta\rho_{\rm n}/\pi f_0}\,$, where $\rho_{\rm
n}$ is the density of the normal component. In view of the
systematic difference between the fit and the measurements in the
normal phase,  the extrapolation to $T < T_\lambda$ is very
reasonable.

\begin{figure}[tb]
\centerline{\includegraphics[width=0.73\linewidth]{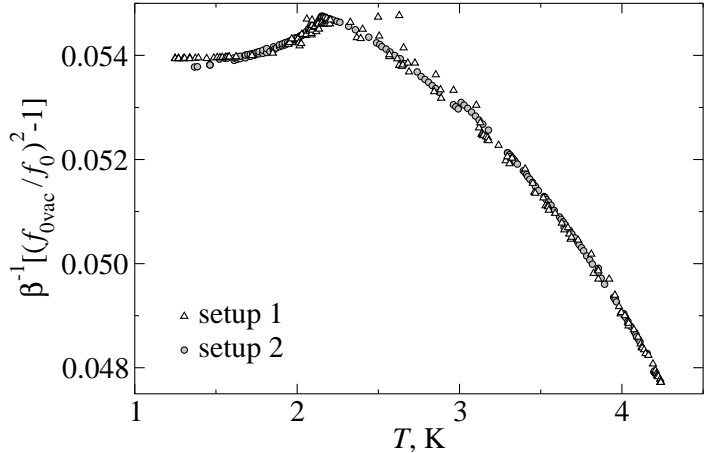}}
\caption{Scaled resonance frequencies of two quartz tuning forks
in liquid $^4$He at saturated vapor pressure. Plotted in this way
both sets of data from Fig.~\ref{he4freq} coincide. The quantity
on the vertical axis is $[(f_{0\rm vac}/f_0)^2 - 1]/\beta $, which
can be considered an ``effective'' value of the density $
\rho/\rho_{\rm q}$. } \label{he4scaled}
\end{figure}

In Fig.~\ref{he4scaled} we plot the same data once more with
$[(f_{0\rm vac}/f_0)^2 - 1]/\beta$ on the vertical scale versus
temperature. Now the two sets of data are reduced on the same
temperature dependence, which is mainly that of the total liquid
density $\rho$. The nice agreement in this plot lends support to
the model expressed by Eq.~(\ref{f0}) and to the extracted values
of $\beta$ for the two forks, derived by fitting their data
separately to Eq.~(\ref{f0}) at temperatures $T > T_\lambda$.
Similar to the $^3$He results in the previous section, we obtain
rather different values for the $\beta$ factors of the two forks
($\beta = 1.39$ and 1.27, Fig.~\ref{he4freq}). These
$\beta$ values are larger than those from the $^3$He measurements,
presumably owing to the can around the fork. Interestingly in
Fig.~\ref{he4scaled}, the resonant frequency, when scaled with the
relevant $\beta$ value, looks promising as a calibration for
thermometry in $^4$He at $T > 1.5\,$K.

\begin{figure}[tb]
\centerline{\includegraphics[width=0.7\linewidth]{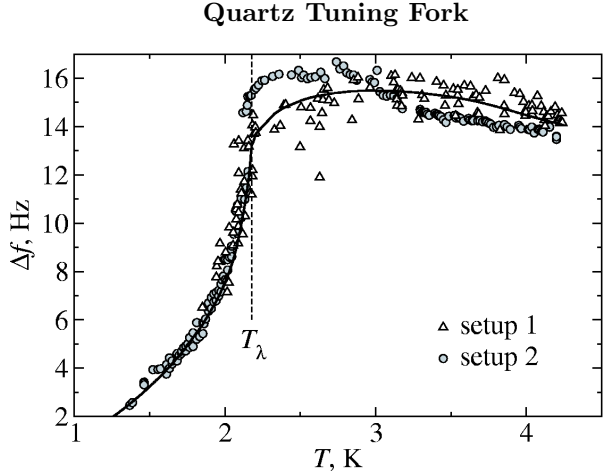}}
\caption{Resonance widths of two quartz tuning forks in liquid
$^4$He at saturated vapor pressure. The symbols represent the
experimental data and the line is a fit of the data from setup 1
at $T>T_\lambda$ to Eq.~(\ref{df}) using $\mathcal{C}$ as fitting
parameter. The fit gives $\mathcal{C} = 0.90$. The fit to the
setup 2 data is almost identical to the one shown and gives
$\mathcal{C} = 0.88$. The data on the physical properties of
liquid $^4$He are from Ref.~[\onlinecite{Russ}].} \label{he4width}
\end{figure}

In Fig.~\ref{he4width} the resonance widths of the two tuning
forks are shown. Above $T_\lambda$ the viscosity of liquid $^4$He
is not a strong function of temperature and a fit of the data to
Eq.~(\ref{df}), using $\mathcal{C}$ as a fitting parameter, works
reasonably well. Thus it is interesting to compare the
extrapolation of the fit to the experimental data in the
temperature regime $T < T_\lambda$. In the extrapolation we use
the same model of non-interacting normal and superfluid
components. The resonance width is only associated with the normal
component and thus in Eq.~(\ref{df}) we replace $\rho$ with
$\rho_{\rm n}$. Remarkably, the two forks follow nicely the
extrapolated dependence in the superfluid regime, in one case down
to 1.8\,K and in the other to 1.4\,K.

An intriguing question is whether any contribution in these
results can be attributed to vortex generation and mutual
friction.\cite{Donnelly} From earlier measurements with vibrating
wires, grids, and spheres it is known that extra damping occurs in
superfluid $^4$He even at low drive from the interaction of the
normal and superfluid components in the presence of quantized
vortices, from mutual friction losses. In $^4$He vortices are
easily pinned on surfaces as the vortex core is of atomic size and
any surface becomes sufficiently rough for pinning. Thus the
surface of the vibrating sensor might be loaded with pinned
remnant vortices. These vortices might originate from thermal
counterflow produced in a rapid cool down through $T_{\rm c}$,
from some other source of residual flow in the system, or if the
oscillating object has been driven previously in the superfluid
state at velocities above some critical value.

In the present measurements shown in Fig.~\ref{he4width}, the
amplitude of the fork current was always kept low (typically below
120\,nA). This current corresponds to velocities less than
1.5\,cm/s, which is below the typical critical velocity in $^4$He
of order 4 -- 5\,cm/s. Additional tests were performed when the
forks were driven hard enough so that non-linear response could be
seen and vortices should have been created. After the drive was
reduced back to low level the width also returned to the original
value shown in Fig.~\ref{he4width} without hysteresis. This
stability in the response is quite unlike the usually observed
differences between ``virgin'' and ``trained'' responses of a
vibrating grid, for instance.\cite{MF_Losses}  If this stability
of the forks can be reliably reproduced and especially if it
persists down to lower temperatures, then the quartz tuning fork
might become a useful thermometer for superfluid $^4$He which is
not troubled by ``vortex layer'' problems. \vspace{-3mm}

\section{CONCLUSIONS}

The quartz tuning fork is a robust and easy-to-use sensor in
cryogenic environments. In view of the results in
Fig.~\ref{he3allT}, it appears to be a useful secondary
thermometer for superfluid $^3$He research. It requires less work
and know how to implement and to operate than any other of the
currently available methods in the superfluid $^3$He temperature
regime. Its response is well described at low excitation in the
linear drive regime in terms of our hydrodynamic model which
includes fitting parameters with a physical origin. More
statistics on different forks are needed to decide whether simple
means can be worked out to fix the parameter values and the
calibration of the device as a thermometer. For such tests the
forks should be adequately thermally cycled and preselected based
on their LHe temperature resonance widths $\Delta f_{\rm vac}$ in
vacuum. The important physics, which should be explored with the
quartz tuning fork, lies at the lowest temperatures in the
ballistic regime, both in $^4$He-II and $^3$He-B, where the
interaction of the fork with quantized vortices should be
investigated. \vspace{-3mm}

\section*{EPILOGUE: Dedication to Frank Pobell}

Oscillating devices immersed in a bath of liquid He were an
important element in Frank Pobell's research. He is remembered for
his passionate mission to reach ever lower temperature records. A
controversial element in this quest was thermometry. The vibrating
wire resonator was and still is the best thermometer for the $T
\rightarrow 0$ limit in $^3$He-B. To explore the limits of this
device, he studied in many papers the nonlinear response and
dissipation in different wire materials down to below 1\,mK in
vacuum. He discovered that even metals display slow heat release
from defect structures, which relax similar to the tunneling model
of two-level systems in glassy amorphous materials.\cite{pobell}
These measurements should now be repeated for the quartz tuning
fork.

The senior members among the authors of this report remember Frank
from this time as an extremely focused and industrious researcher.
He was a visitor in the Low Temperature Laboratory for two months
during the spring term of 1991. One of us (MK) was sharing the
office room with him. He was working from early morning until late
evening without break on his administrative chores, writing his
research reports, and examining the manuscripts submitted for
publication in JLTP. We were so impressed by this diligence and
the steady flow of new results. An excellent example is his
book\cite{Pobell} "Matter and Methods at Low Temperatures", which
he had just completed and which appeared later in the same year.
This book is still gratefully used as the best text book for our
courses in low temperature physics. East of Germany Frank is
remembered (LS and PS) for his help and support to the low
temperature community behind the iron curtain before it was lifted
in 1989. This special relationship has continued over the years,
an example was Frank's co-chairmanship of the LT21 Conference in
1996 in Prague.

This research is supported in part by the EU research program
ULTI-4 (RITA-CT-2003-505313), by the Institutional Research Plan
AVOZ10100520  and the Czech Grant Agency under grant
GACR202/05/0218, by the Slovak Grant Agency VEGA under grant
2/6168/06 and by the SAS under grant CE I-2/2003.


\begin{thebibliography}{99}
\bibitem{forks-he3} The fork used in the measurements at 29\,bar pressure
  was manufactured in Japan and was obtained from RS Company under
  stock number 472-1161, brand Fox Electronics and manufacturer's
  part number NC26. The fork used in the zero pressure
  measurements was manufactured in Taiwan and was obtained from RS
  Company under stock number 226-1437, brand C-MAC Frequency
  Products and manufacturer's part number A103C.
\bibitem{forks-he4} In setup 1 the same type of fork was used as in
  $^3$He measurements at 29\,bar pressure.\cite{forks-he3} In
  setup 2 the fork was made in China by Fronter Electronics,
  product number DT26.
\end{thebibliography}
\end{document}